\documentclass[aps,prl,reprint,showpacs,floatfix]{revtex4-1}

\usepackage{color}

\usepackage[utf8]{inputenc}
\usepackage{amsmath,amssymb,amstext}
\usepackage{amsthm}
\usepackage{dsfont}
\usepackage{graphicx}

\def\sout{\bgroup\markoverwith
{\textcolor{red}{\rule[0.5ex]{2pt}{0.5pt}}}\ULon}
\def\be{\begin{equation}}
\def\ee{\end{equation}}
\def\bes{\begin{equation*}}
\def\ees{\end{equation*}}
\def\bea{\begin{eqnarray}}
\def\eea{\end{eqnarray}}
\def\beas{\begin{eqnarray*}}
\def\eeas{\end{eqnarray*}}
\def\bal#1\eal{\begin{align}#1\end{align}}
\def\bals#1\eals{\begin{align*}#1\end{align*}}

\newcommand{\ket}[1]{|#1\rangle}

\usepackage[normalem]{ulem}

\usepackage{braket}

\usepackage{pdfpages}
\usepackage{pgffor}
\makeatletter
\AtBeginDocument{\let\LS@rot\@undefined}
\makeatother

\begin{document}

\title{Analytical approach to the Bose polaron \\ via $q$-deformed Lie algebra}


\author{E. Yakaboylu}
\email{enderalp.yakaboylu@mpq.mpg.de}
\affiliation{Max Planck Institute of Quantum Optics, 85748 Garching, Germany}

\date{\today}

\begin{abstract}

We present a novel approach to the Bose polaron based on the notion of quantum groups, also known as $q$-deformed Lie algebras. In this approach, a mobile impurity can be depicted as a deformation of the Lie algebra of the bosonic creation and annihilation operators of the bath, in which the impurity is immersed. Accordingly, the Bose polaron can be described as a bath of noninteracting $q$-deformed bosons, which allows us to provide an analytical formulation of the Bose polaron at arbitrary couplings. Particularly, we derive its ground state energy in the phonon branch of the Bogoliubov dispersion and demonstrate that the previously observed transition from a repulsive to an attractive polaron occurs at the vicinity where the quantum group symmetry is broken. Furthermore, our approach has the potential to open up new avenues in polaron physics by connecting it with seemingly unrelated research topics where quantum groups play an essential role, such as anyons.

\end{abstract}

\maketitle

The polaron model, which was initially introduced for a quasiparticle consisting of an electron dressed by lattice excitations in a crystal~\cite{landau1933uber,pekar1946local,frohlich1954electrons}, has been adapted to several different contexts, and nowadays is used to describe more general classes of impurities interacting with a quantum many-particle bath. In particular, within recent advances in ultracold atomic physics, Bose polarons, which correspond to quantum impurities interacting with a surrounding Bose-Einstein condensate (BEC), broke new ground in the study of polaron physics. Although, Bose polarons have been investigated intensively in several experimental~\cite{Schirotzek_09,Jorgensen_16,Hu_2016,
yan2020bose,Camargo_18,koepsell2019imaging} and theoretical~\cite{grusdt2015new,Shchadilova_2016,Grusdt_2017,
grusdt2017bose,Volosniev_2017,Kain_2018,Yoshida_2018,Levinsen_2021} studies, there remain still many open questions, which shape the current research directions in the field. Probably the first one coming to mind is whether it is possible to provide a complete analytical description of Bose polarons at arbitrary couplings. In the present Letter, inspired by the phenomenon of how a single impurity drastically changes its surrounding bath~\cite{kondo1964resistance,Anderson_67}, we address the aforementioned question within the concept of quantum groups.

Quantum groups, also known as $q$-deformed Lie algebras~\cite{drinfeld1987proceedings,
jimbo1986q,jimbo1990yang,kulish1981quantum}, are Hopf algebras possessing a coproduct, a counit, and an antipode, in addition to the regular structures of an algebra~\cite{abe1977hopf}. In physics, quantum groups have been first applied to solve the quantum Yang-Baxter equation~\cite{jimbo1990yang}. Although quantum group symmetries are believed not to be common in physical problems, over the years, they have found several applications to spin chains~\cite{batchelor1990q}, anyons~\cite{lerda1993anyons,ubriaco1997anyonic}, quantum
optics~\cite{buvzek1992jaynes,chaichian1990quantum}, nuclear physics~\cite{bonatsos1999quantum}, rotational and vibrational molecular spectra~\cite{esteve1992q,raychev1995quantum}, and  molecular impurities~\cite{Yakaboylu_2018_quantum}. 

Now, we demonstrate that the Bose polaron features quantum group symmetries within the framework of the extended Bogoliubov-Fr\"{o}hlich Hamiltonian, which provides the most accurate model for the Bose polaron as long as the Bogoliubov description of the Bose gas is valid~\cite{grusdt2017bose}. Particularly, we show that the Lie algebra of the bosonic creation and annihilation operators of the bath, called the Heisenberg-Weyl algebra, can be deformed due to the presence of an impurity. Such an approach, first, allows us to introduce a simple parent Hamiltonian in terms of $q$-deformed bosons, whose series expansion in the usual bosons corresponds to the extended Bogoliubov-Fr\"{o}hlich Hamiltonian in the branch where the Bogoliubov dispersion takes the phonon-like linear form. Furthermore, as an insightful physical picture, the parent Hamiltonian describes the gas of interacting Bogoliubov bosons in terms of a gas of noninteracting $q$-deformed bosons. Within this picture, we derive the ground state energy of the Bose polaron analytically in the phonon branch and demonstrate that at the vicinity of a critical value of the $q$-parameter, where the quantum group symmetry is broken, the Bose polaron changes from repulsive to attractive.

We start by considering a single impurity immersed in a weakly interacting Bose gas. Although our approach is independent of dimension, for the sake of simplicity, we will consider the impurity-bath system in one dimension with the quantization length $L$, which we set to unity. Within the Bogoliubov approximation, the most accurate Hamiltonian describing the Bose polaron is given by
\bal
\nonumber H & = \frac{P^2}{2} + \sum_k \omega_k b^\dagger_k b_k + \sum_k \lambda_k \, e^{-i k X} \left( b^\dagger_k + b_{-k} \right) \\
& + \sum_{k,k'} \lambda^+_{k,k'} \, e^{-i (k-k') X}b^\dagger_k b_{k'} \\
\nonumber & + \frac{1}{2 }  \sum_{k,k'} \lambda^-_{k,k'} \, e^{-i (k+k') X}  \left(b^\dagger_k b^\dagger_{k'} +  b_{-k} b_{-k'} \right) \, ,
\eal
which might be called the extended Bogoliubov-Fr\"{o}hlich Hamiltonian, see Refs.~\cite{Shchadilova_2016,Grusdt_2017,grusdt2017bose,Kain_2018} for its detailed derivation. Here, the first line corresponds to the well-known Fr\"{o}hlich Hamiltonian, where $P^2/2$ is the kinetic energy operator of the impurity with mass $M=1$ and $X$ is the position operator. $b^\dagger_k$ and $b_k$ are the Bogoliubov creation and annihilation operators, $\omega_k$ is the Bogoliubov dispersion relation, $\lambda_k = g_\text{IB} \sqrt{n_\text{B}} \chi_k$ is the linear coupling, $g_\text{IB}$ is the impurity-boson coupling constant, $n_\text{B}$ is the condensate density, and $\chi_k = \sqrt{\varepsilon_k / \omega_k}$, with the bare bosons' energy  $\varepsilon_k = k^2/(2m_\text{B}) $ and mass $m_\text{B}$. The second and third lines, which are beyond the Fr\"{o}hlich paradigm~\cite{grusdt2017bose}, describe two-phonon scattering processes with the quadratic couplings $\lambda^{\pm}_{k,k'} = g_\text{IB} (\chi_k \chi_{k'} \pm \chi_k^{-1} \chi_{k'}^{-1} )/2$.

As the total linear momentum of the system is conserved, one can decouple the impurity's degree of freedom from the bath by applying the Lee-Low-Pines (LLP) transformation~\cite{LLP_53}, $U_\text{LLP} = \exp(- i X \sum_k b^\dagger_k b_k)$, which translates the creation and annihilation operators; $U_\text{LLP}^{-1} b^\dagger_k U_\text{LLP} = \exp(i k X) b^\dagger_k $. The transformed Hamiltonian, $H_\text{B} = U_\text{LLP} ^{-1} \, H \, U_\text{LLP}$, yields
\bal
\label{B_ham}
& \nonumber H_\text{B} = \sum_k \tilde{\omega}_k b^\dagger_k b_k +\sum_k \lambda_k \left( b^\dagger_k +  b_k \right) + \sum_{k,k'} \lambda^+_{k,k'} b^\dagger_k b_{k'} \\
  & + \frac{1}{2}\sum_{k,k'} \lambda^-_{k,k'}  \left(b^\dagger_k b^\dagger_{k'} +  b_{k} b_{k'} \right) + \frac{1}{2}\sum_{k,k'} k k' b^\dagger_k b^\dagger_{k'} b_{k'} b_k ,
\eal
where we replaced the momentum operator $P$ with its eigenvalue $p$, defined $\tilde{\omega}_k = \omega_k - p k + k^2/2$, and omitted the constant term $p^2/2$. 

The Hamiltonian~\eqref{B_ham} is the main focus of the present Letter. We underline that as it is purely bosonic, the Hamiltonian~\eqref{B_ham} describes a bath of interacting Bogoliubov bosons. Furthermore, its spectrum, in principle, can be identified solely by using the Heisenberg-Weyl algebra of the bosonic creation and annhilation operators;
\be
\label{H_W_algebra}
[b_k,b^\dagger_k] = 1 \,, \quad [B_k, b^\dagger_k] = b^\dagger_k \, , \quad [B_k, b_k] = - b_k \, ,
\ee
with $B_k  =  b^\dagger_k b_k $ being the number operator. We will show that the Hamiltonian~\eqref{B_ham} can be written elegantly in terms of the elements of the deformed Heisenberg-Weyl algebra. This allow us to approach the Bose polaron within quantum groups.

In order to present the method in a transparent way, we first focus on the case of a single-phonon-mode and discuss the full Hamiltonian later. In the former case, the Hamiltonian~\eqref{B_ham} can be given by
\bal
\label{B_toy_0}
 H_\text{sm} = \bar{\omega} b^\dagger b + \lambda (b^\dagger + b) + \frac{\lambda^-}{2} (b^\dagger b^\dagger + b b) + \frac{k^2}{2} b^\dagger b^\dagger b b  \, ,
\eal
where we define $\bar{\omega} = \tilde{\omega} + \lambda^+$ and, for a shorthand notation, we omit the index of the particular mode $k$ from the relevant expressions, i.e., $b^\dagger_k \to b^\dagger$, $\bar{\omega}_k \to \bar{\omega}$, and so on. This Hamiltonian, by summing over $k$, can also be used to approximate the full Hamiltonian, when the elastic scattering ($k' = k$) is dominant.
 
We now apply the displacement operator, $U_\text{disp} = \exp(\alpha \, b^\dagger- \alpha^* \, b)$, which shifts the creation and annihilation operators; $U_\text{disp}^{-1} b^\dagger U_\text{disp} = b^\dagger + \alpha^*$ with a complex parameter $\alpha$. The transformed Hamiltonian, $H_\text{sm}' = U_\text{disp}^{-1}  H_\text{sm}  U_\text{disp} $, can be written as
\be
\label{B_toy1}
 H_\text{sm}'   =  v \, b^\dagger \left(1 + \frac{k^2}{2 v} b^\dagger b \right) b  +  u \,  b^\dagger \left(1 + \frac{\alpha k^2}{u} b^\dagger b \right) + \text{h.c.}  \,,
\ee
where $v = \bar{\omega} + 2 |\alpha|^2 k^2$, $u = \lambda + \bar{\omega} \alpha$, $\text{h.c.}$ stands for the Hermitian conjugate of the preceding terms, and we omitted the constant term, $\bar{\omega} |\alpha|^2 + \lambda (\alpha + \alpha^*)  +  \lambda^- (\alpha^2 + \alpha^{* \, 2})/2 + k^2 |\alpha|^4 /2$. In the derivation of the Hamiltonian~\eqref{B_toy1} we eliminated the quadratic term, $(\lambda^- + k^2 \alpha^2)b^\dagger b^\dagger/2 + \text{h.c.}$, by choosing $ \alpha^2  = \alpha^{* \, 2} = - \lambda^- /k^2 $. For the sake of simplicity, we further assume $\lambda^- < 0$ and hence $\alpha \in \mathbb{R}$.

The structure of the Hamiltonian~\eqref{B_toy1} inspires us to introduce deformed creation and annihilation operators that can simplify the Hamiltonian further. Particularly, from the first term, we define a deformed operator, whose series expansion is in the form of
\be
\label{deform_form}
a^\dagger = b^\dagger \left( 1 + \rho \, b^\dagger b + \cdots \right) \, ,
\ee
with $\rho = \sqrt{1 + k^2/(2 v)} - 1$, such that the series expansion of the term, $v\, a^\dagger a$, truncated at the quartic order in $b^{(\dagger)}$ corresponds to the first term of the Hamiltonian~\eqref{B_toy1}. The second term of the Hamiltonian also matches with the expansion of the term, $u\, (a^\dagger + a) $, if there exists the relation, $\sqrt{1 + k^2/(2 v)} - 1 = \alpha k^2 / u $. This relation, which can be interpreted as an expression of the quantum group symmetry, imposes certain restrictions on the values of the dispersion relation as well as the coupling. However, as we will discuss later, the physical parameters of the BEC, indeed, meet such a relation for the full Hamiltonian. Therefore, for the single-phonon-mode case, without loss of physical generality, we assume the existence of this symmetry. As a matter of fact, one can always consider the certain phonon mode, $k$, which satisfies the relation.

The Hamiltonian~\eqref{B_toy1}, then, can be identified as the quartic order series expansion in the usual creation and annihilation operators of the following deformed parent Hamiltonian
\be
\label{ham_spm_q}
H_\text{sm}^\text{def} = v \, a^\dagger a + u (a^\dagger + a) \, ,
\ee
where $v$ can be interpreted as the dispersion relation of the deformed bosons. As we discuss below, the expansion can be truncated, if the higher order terms can be neglected. This allows us to derive the spectrum of the Hamiltonian~\eqref{B_toy1} by using the parent Hamiltonian~\eqref{ham_spm_q}.

The Hamiltonian~\eqref{ham_spm_q} is simply in the form of a shifted harmonic oscillator Hamiltonian describing noninteracting deformed bosons. Therefore, the system of interacting regular bosons, which is originally governed by the Hamiltonian~\eqref{B_toy_0}, can be identified as a system of free deformed bosons. This follows from the fact that the introduced deformed bosons accumulate certain correlations which describe implicit interactions between the ordinary bosons. This is in analogy to the Bogoliubov transformation, which describes an interacting Bose gas in terms of a gas of noninteracting quasiparticles~\cite{Pitaevskii2016}. However, there is a price we pay. Despite of its simple form, finding the spectrum of the Hamiltonian~\eqref{ham_spm_q} is not trivial due to the deformed commutation relations, which we define now.

\begin{figure}[t]
\centering
\includegraphics[width=0.9\linewidth]{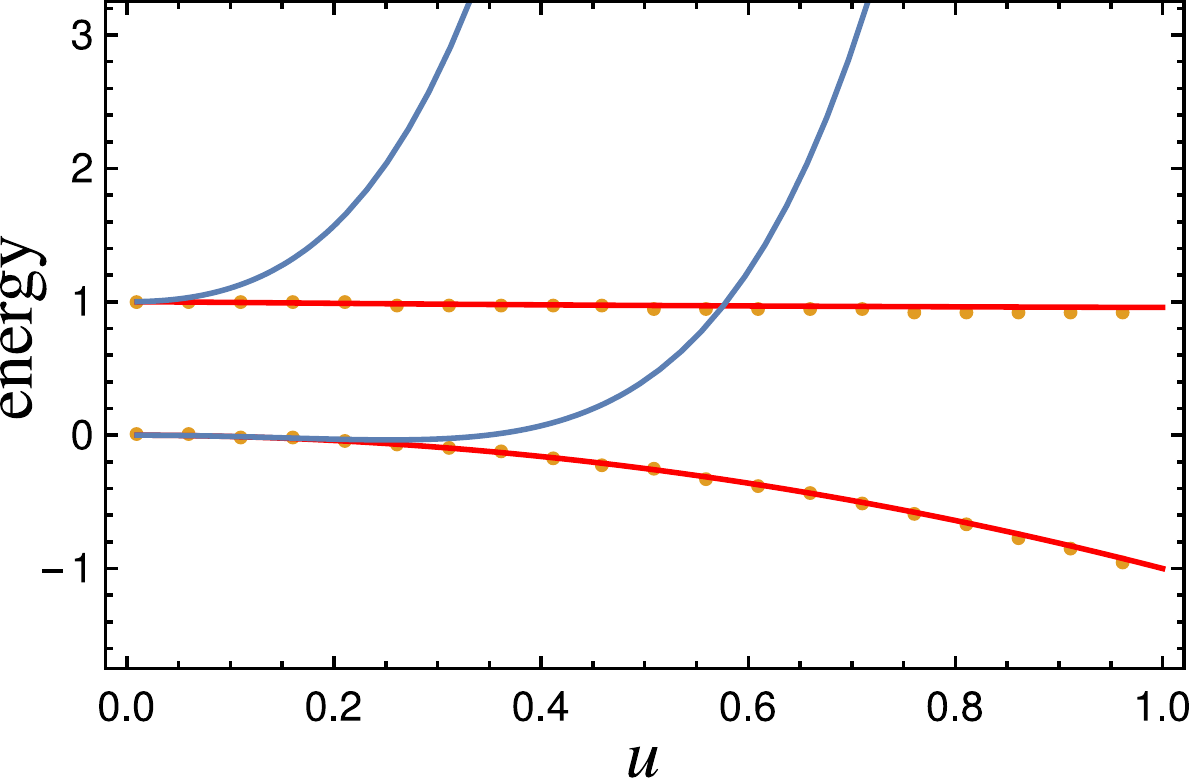}
\caption{Comparison of the first two eigenvalues obtained from a numerical diagonalization of the Hamiltonian~\eqref{B_toy1} (yellow dotted) and the analytical result~\eqref{q_energy} (red) as a function of the effective coupling $u$. We further plot the perturbative result (blue). The parameters are chosen as $v=1$ and $\alpha = - 0.2$.}
\label{fig1}
\end{figure}

The form of the deformed operator~\eqref{deform_form} is known in the context of $q$-deformation, see for instance Ref.~\cite{raychev1995quantum}. It is an element of the called $q$-deformed Heisenberg-Weyl algebra~\cite{arik1976hilbert}. This deformed algebra is defined by the group elements $a^\dagger$, $a$, and $N$, which satisfy the following commutation relations
\be
\label{q_hw_a}
a a^\dagger - q a^\dagger a = 1 \, , \quad [N, a^\dagger] = a^\dagger \, , \quad [N, a]= -a \, ,
\ee
with $q$ being the deformation parameter. We note that the number operator $N$ does not equal to $a^\dagger a$. It is also worth noticing that the $q$-commutation relation in Eq.~\eqref{q_hw_a} can be used to build anyonic algebras~\cite{lerda1993anyons,ubriaco1997anyonic}.

One particular realization of this algebra can be fulfilled with the following expressions
\be
\label{realization}
a^\dagger = \sqrt{\frac{[N]}{N}} b^\dagger \, , \quad a = b \sqrt{\frac{[N]}{N}} \, , \quad N = B = b^\dagger b \, .
\ee
Here, we recall that $b$ and $b^\dagger$ are the usual bosonic operators, and $[N] = (q^N - 1)/(q-1)$. In the limit of $q \to 1$, one recovers the usual Heisenberg-Weyl algebra~\eqref{H_W_algebra}. One can obtain the expressions for $a^\dagger $ and $a$ in the form of infinite series of $B$; $a^\dagger = \sum_{n=0} c_n B^n b^\dagger$, which allows us to define the $q$-deformed bosons in terms of the usual ones. Specifically, by using the identity $(b^\dagger b)^l = \sum_{k=1}^l S(l,k) (b^\dagger)^k b^k$, with  $S(l,k)$ being Stirling numbers of the second kind, the $q$-deformed creation operator can be written as
\be
\label{q_deform_hw}
a^\dagger = b^\dagger \left(1 + \left(\sqrt{[2]/2} -1 \right) b^\dagger b + \cdots \right)  \, .
\ee
This series is in the form of Eq.~\eqref{deform_form} with $\rho = \sqrt{[2]/2} -1 $, which identifies the deformation parameter as $q=1 + k^2 /v$.  As the deformation part, $k^2/v$, results from the LLP transformation, we interpret that the Lie algebra of the bath is deformed by the impurity. Moreover, we deduce from the series expansion~\eqref{q_deform_hw} that the Hamiltonian~\eqref{ham_spm_q} can be truncated at the quartic order under the condition of $\zeta \ll 1$, see~\footnote{Supplement Material} for the definition of $\zeta$.

In order to find the spectrum, as an initial guess, one might think of applying the displacement operator to shift the $q$-deformed creation and annihilation operators. However, as we mentioned earlier, due to the non-trivial commutation relation, this is not possible. In fact, it is known that there does not exist a unitary displacement operator for the $q$-deformed Weyl-Heisenberg algebra~\cite{arik1976hilbert,mcdermott1994analogue,aneva2005deformed}. Nevertheless, we find the eigenvalues exactly
\be
\label{q_energy}
E_\text{sm}^{\text{def} \, (n)} = v [n] - \frac{1}{q^n}\frac{u^2}{v} \, ,
\ee
see~[34] for the derivation. In the limit of $q \to 1$, one recovers the well-known eigenvalues; $E_\text{sm}^{\text{def} \, (n)} \to v n - u^2/v$. We would like to emphasize that solving the eigenvalue equation of the Hamiltonian~\eqref{ham_spm_q} is, in itself, a very important result for the field of quantum groups, as it corresponds to the resolution of the problem of the $q$-displacement operator.

In summary, under the condition of $\zeta \ll 1$, the ground state energy of the single-phonon-mode Hamiltonian~\eqref{B_toy1} coincides with that of the deformed parent Hamiltonian, and hence given by Eq.~\eqref{q_energy}. This correspondence turns the picture of interacting regular bosons into the one of  noninteracting $q$-deformed bosons. In order to illustrate the ability of the analytical result~\eqref{q_energy}, in Fig.~\ref{fig1}, we compare it with the exact numerical result, which we calculated by diagonalizing the Hamiltonian~\eqref{B_toy1} in the basis of the eigenstates of the number operator, $\ket{m}$. Instead of the original parameters appeared in the Hamiltonian~\eqref{B_toy_0}, the eigenvalues are calculated in terms of the parameters, $v$, $u$, and $\alpha$, of the Hamiltonian~\eqref{B_toy1}, which is unitarily equivalent to the former. In the figure, the parameter $\alpha$ is chosen to satisfy the condition $\zeta \ll 1$. In the same figure, we also present the result of the first order perturbation theory, where the term, $k^2 b^\dagger b^\dagger b b/2 + \alpha k^2 b^\dagger b^\dagger b + \text{h.c.}$, is treated perturbatively. It is clear that the analytical result agrees with the numerical one even beyond the perturbative regime.

Now we will focus on the full Hamiltonian~\eqref{B_ham}. In a similar way, in order to eliminate the quadratic terms we apply the displacement operator, $U_\text{disp} = e^{\sum_k (\alpha_k b^\dagger_k - \alpha_k^* b_k) }$. The transformed Bogoliubov-Fr\"{o}hlich Hamiltonian can be written as
\bal
\label{B_ham_1}
 H'_\text{B} & = \sum_k v_k \, b^\dagger_k \left( 1 + \sum_{k'} \frac{k k'}{2 v_k} b^\dagger_{k'} b_{k'} \right) b_k \\
\nonumber & + \sum_k u_k \, b^\dagger_k \left( 1 + \sum_{k'} \frac{ k k' \alpha_k}{u_k} b^\dagger_{k'} b_{k'} \right) + \text{h.c.}  \, ,
\eal
where $v_k = \tilde{\omega}_k + k \sum_{k'} k' |\alpha_{k'}|^2$, $u_k = \lambda_k + \alpha_k ( \tilde{\omega}_k -  k \sum_{k'} k' |\alpha_{k'}|^2) $, and the constant term, $ \sum_k \tilde{\omega}_k |\alpha_k|^2  -3 \sum_{k,k'} k k' |\alpha_k|^2 |\alpha_{k'}|^2 /2$, was omitted. In the derivation of Eq.~\eqref{B_ham_1}, the quadratic terms $b^\dagger_k b^\dagger_{k'}$, $ b_k b_{k'}$, and $b^\dagger_k b_{k'}$ are canceled out by specifying $\alpha_k \alpha_{k'} = - \lambda^-_{k,k'}/(k k')$ and $\alpha_k \alpha_{k'}^* = - \lambda^+_{k,k'}/(k k')$. This is possible only if the couplings satisfy the relation, $|\lambda^+_{k,k'}| = |\lambda^-_{k,k'}|$. Strikingly, the relation is satisfied via $\lambda^+_{k,k'} = - \lambda^-_{k,k'} \approx g_\text{IB} \chi_k^{-1} \chi_{k'}^{-1} /2 $ in the phonon branch of the Bogoliubov dispersion, i.e., when $\omega_k \approx c k $ with $c = \sqrt{g_\text{BB} n_\text{B}/m_\text{B}}$ being the sound velocity
and $g_\text{BB}$ the interaction strength between bosons, see Ref.~\cite{Pitaevskii2016}. The relation, $\lambda^+_{k,k'} = - \lambda^-_{k,k'} $, further implies that $\alpha_k = i |\alpha_k|$, and hence, allows us to consider the non-trivial phases of $g_\text{IB} < 0$, where the polaron changes from repulsive to attractive as $g_\text{IB}$ increases across a critical value~\cite{Rath_2013,Kain_2018}.

\begin{figure}[t]
\centering
\includegraphics[width=0.9\linewidth]{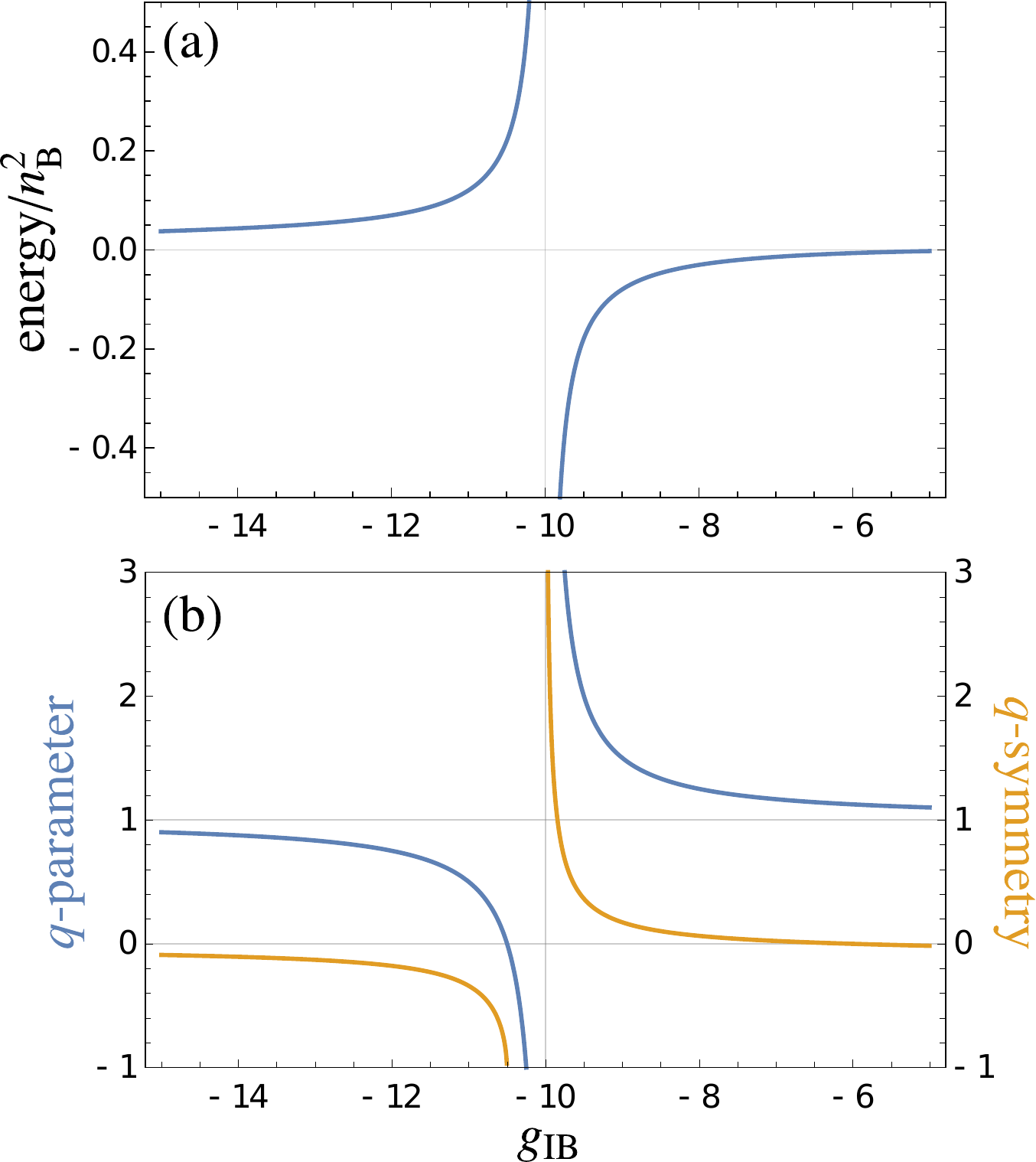}
\caption{(a) The ground state energy of the Bose polaron, given by Eq.~\eqref{ground_state_B}, as a function of $g_\text{BI}$. The Bose polaron undergoes a transition from repulsive to attractive as $g_\text{BI}$ increases. (b) The deformation parameter $q(k,k')$ (blue) and the parameter expressing the quantum group symmetry, $\rho (k,k') - k k' \alpha_k / u_k$, (yellow), which are calculated at $k=k'=\Lambda/10$. The transition is located at the vicinity where the quantum group symmetry is broken.  The applied parameter are $\Lambda = c$, $g_\text{BB} = 1$, $c = 10$, $m_\text{B} = 1$, and $p=0$.}
\label{fig2}
\end{figure}

The form of the Hamiltonian~\eqref{B_ham_1}, now, allows us to introduce deformed operators. For a later purpose, we first symmetrize the term $k k' / (2 v_k)$ by replacing it with $k k' / (v_k + v_{k'})$ in the first line of the Hamiltonian~\eqref{B_ham_1}. This replacement has no effect, as the quartic term, $\sum_{k ,k'} k k' b^\dagger_k b^\dagger_{k'} b_{k'} b_k / 2$, is symmetric with respect to the exchange of $k$ and $k'$, but the remainder term is anti-symmetric. Then, we introduce the following deformed operator
\be
\label{deform_a_full}
a^\dagger_k = b^\dagger_k \, \prod_{k'} q_{k k'}^{B_{k'} /2} = b^\dagger_k \left( 1 + \sum_{k'} \rho (k,k') b^\dagger_{k'} b_{k'} + \cdots \right).
\ee
Here, the deformation parameter, $q_{kk'} = (1 + \rho (k,k'))^2$, is a complex number as the coupling $u_k$ is, and we define $\rho (k,k')$ via the relation, $\rho (k,k') + \rho^* (k,k') + |\rho (k,k')|^2 = k k'/(v_k+ v_{k'})$, such that it corresponds to the first line of the Hamiltonian~\eqref{B_ham_1}. We emphasize that in the phonon branch of the Bogoliubov dispersion, where $k \ll m_\text{B} c$, the expansion~\eqref{deform_a_full} naturally truncates at the quartic order, as the higher orders are proportional to $k / (m_\text{B} c)$. Furthermore, the relation, $\rho (k,k') \approx k k' \alpha_k / u_k$, also holds true, see Fig.~\eqref{fig2}~(b), thereby the deformed operator~\eqref{deform_a_full} matches with the second line of the Hamiltonian~\eqref{B_ham_1}, as well. This is basically the manifestation of the quantum group symmetry of the Bose polaron. Thus, the full Bogoliubov-Fr\"{o}hlich Hamiltonian~\eqref{B_ham_1} in the phonon branch can be written as the following deformed parent Hamiltonian
\be
\label{q_full}
H_\text{B}^\text{def} = \sum_k v_k a^\dagger_k a_k + \sum_k ( u_k \, a^\dagger_k + u^*_k \, a_k) \, .
\ee

The Hamiltonian~\eqref{q_full} is elegantly in the form of a collection of uncoupled shifted harmonic oscillator so that it describes a gas of noninteracting deformed bosons. However, as a price to pay, one has to deal with the complicated deformed Lie algebra, which can be written as
\bal
\label{q_hw_a_full}
& a_k a_{k}^\dagger - |q_{kk}| \, a_{k}^\dagger a_k = \prod_{k'} |q_{kk'}|^{B_{k'}} \, , \\
\nonumber & [N_k, a_{k}^\dagger] = a_{k}^\dagger\, , \quad [N_k, a_{k}]= -a_{k} \, ,
\eal
with $N_k = B_k$. Moreover, in contrast to the case of the usual algebra, the commutation relation between different modes ($k\neq k'$) do not vanish, but are given by
\begin{subequations}
\bal
 & a_k a_{k'}^\dagger - \sqrt{q^*_{kk'}}\sqrt{q_{k'k}} \, a_{k'}^\dagger a_k = 0 \, ,\\
\label{def_com_aa} & \sqrt{q^*_{k'k}}\, a_k a_{k'}- \sqrt{q^*_{kk'}} \, a_{k'} a_k = 0 \, .
\eal
\end{subequations}
These commutators lead to further complications for finding the energy eigenvalues. Nevertheless, since the deformation parameter is symmetric, i.e., $q_{kk'} = q_{k'k}$, the commutator~\eqref{def_com_aa} simple reads $[a_k, a_{k'}] = 0 $. This simplification allows us to construct the ground state in terms of the eigenstate of the annihilation operator $a_k$, and hence to derive the ground state energy of the deformed parent Hamiltonian~\eqref{q_full} as
\be
\label{ground_state_B}
E_\text{B}^{\text{def} \, (0)} = - \sum_{k} \frac{|u_k|^2}{v_k} \, ,
\ee
which, consequently, identifies also the ground state energy of the full Bogoliubov-Fr\"{o}hlich Hamiltonian in the phonon branch.

Before presenting a numerical result, we would like to further note the following. It is known that the Bose polaron shows an infrared (IR) divergent behavior in the phonon branch, which requires certain regularization techniques, see, for instance, Ref.~\cite{grusdt2017bose}. The explicit form of the ground state energy~\eqref{ground_state_B} enables us to regularize the IR divergence straightforwardly by means of the Hadamard fine part method. For instance, by setting $p=0$ and $m_\text{B} = 1$ we present a simple and explicit formula for the ground state energy of the Bose polaron:
\be
\label{explicit_formula}
E_\text{B}^{\text{def} \, (0)} = - g_\text{IB} c^2 \frac{(1 - g_\text{IB}/\Lambda)^2}{\Lambda + g_\text{IB}} -  \frac{g_\text{IB}^2 n_\text{B}  \, \Lambda^2}{2c^2 (\Lambda + g_\text{IB})} \, ,
\ee
where $\Lambda$ is the momentum cutoff, until which the Bogoliubov dispersion is linear, and it can be identified via the healing length; $\Lambda \sim 1/ \xi$.

The formula~\eqref{explicit_formula} shows explicitly that there exists a transition from a repulsive to an attractive polaron at the critical value of $g_\text{IB} = - \Lambda$, which is demonstrated in Fig.~\ref{fig2}~(a). This is qualitatively consistent with the previously reported results~\cite{Shchadilova_2016,Grusdt_2017,grusdt2017bose,Kain_2018}. The divergent nature of the energy observed in the figure is a direct consequence of the broken quantum group symmetry at the vicinity of the critical value of $g_\text{IB}$, where the value of the parameter $\rho(k,k')$ deviates from the value of $k k' \alpha_k/u_k$, see the yellow curve in Fig.~\ref{fig2}~(b). In fact, in this regime $\rho(k,k') \propto (1 + g_\text{IB}/\Lambda)^{-1/2}$ and the bosonic expansion of the Hamiltonian~\eqref{q_full} cannot be truncated at the quartic order. Thereby, this is the regime where the picture of the $q$-deformed bosons breaks down. This is in analogous to the called quasiparticle instability, at which the quasiparticle picture fails~\cite{Lemeshko_2015}. Moreover, as shown in Fig.~\ref{fig2}~(b), the behavior of the deformation parameter, $q(k,k')$, changes as the coupling increases across the critical value. While the deformation is negative for the repulsive polaron, it becomes positive for the attractive one (blue curve). We also observe in the same figure that far from the critical value, the deformed bosons coincide with the usual Bogoliubov bosons, as $q(k,k') \to 1$, so that the quantum group symmetry becomes trivial.

In conclusion, in the phonon branch of the Bogoliubov dispersion, where $k \ll m_\text{B} c$, the extended Bogoliubov-Fr\"{o}hlich Hamiltonian~\eqref{B_ham} can be written in terms of noninteracting $q$-deformed bosons, which allow us to derive the ground state energy of the Bose polaron analytically. Our result can be straightforwardly extended to three dimensions and, by including the omitted constant terms, the analytical result can be compared more quantitatively with an \textit{ab initio} calculation. Lastly, as our approach has connected seemingly unrelated branches of many-body physics and mathematics, we anticipate that the Bose polaron and, in general, quantum impurities can be explored in different contexts where quantum groups play crucial roles. Particularly, the deformation of the bosonic commutation relations due to the presence of an impurity might open new horizons for anyons, the basis of topological quantum computing.

\begin{acknowledgments}

The author is grateful to Areg Ghazaryan, Mikhail Lemeshko, Douglas Lundholm, Richard Schmidt, Robert Seiringer, Engin Torun, and Artem Volosniev for valuable discussions, and thanks T\"{u}rk\^an Kobak for assisting with the numerical analysis.

\end{acknowledgments}

\bibliography{anyon_ref.bib}

\newpage

\foreach \x in {1}
{%
\clearpage
\includepdf[pages={\x}]{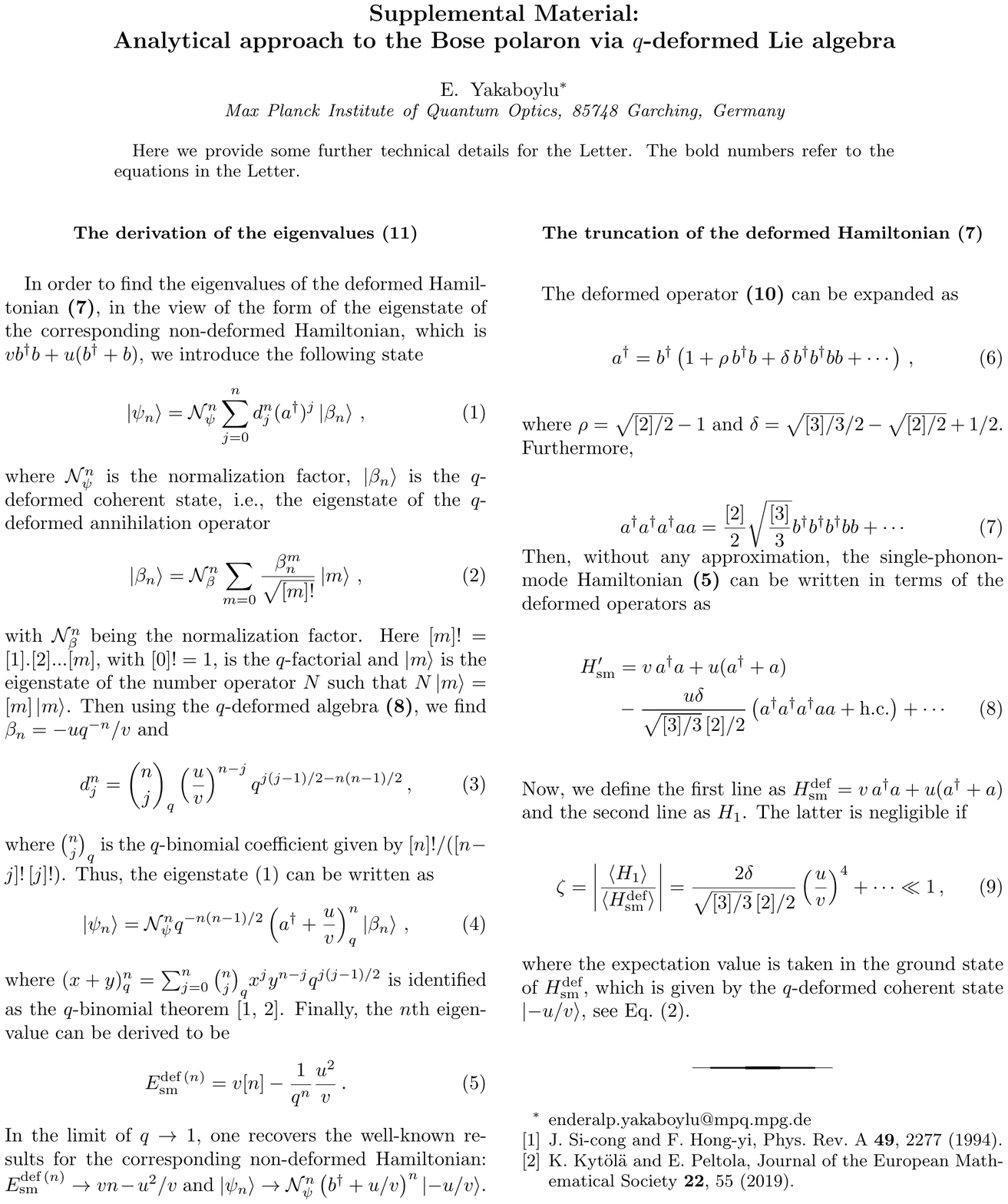} 
}

\end{document}